\documentclass[aps,reprint,amsmath,amssymb,superscriptaddress,prl,floatfix]{revtex4-1}
\usepackage[table,xcdraw]{xcolor}
\usepackage{graphicx}
\usepackage{bm}
\usepackage[colorlinks=true,citecolor=blue,linkcolor=blue,anchorcolor=blue,urlcolor=blue]{hyperref}
\usepackage{mathrsfs}
\usepackage[utf8]{inputenc}
\usepackage{tensor}
\usepackage{physics}
\allowdisplaybreaks[4]
\usepackage[T1]{fontenc}

\newcommand{\tcb}{\textcolor{blue}}

\begin{document}

\title{Probing holographic conformal field theories}
\author{Ming Zhang}
\email{mingchang@outlook.com}
\affiliation{School of Physics, Jiangxi Normal University, Nanchang 330022, China}
\affiliation{Department of Physics and Astronomy, University of Waterloo,  200 University Ave W, Waterloo, Ontario N2L 3G1, Canada}
\affiliation{Perimeter Institute for Theoretical Physics, 31 Caroline St. N., Waterloo, Ontario N2L 2Y5, Canada}
\author{Jiayue Yang}
\email{j43yang@uwaterloo.ca}
\affiliation{Department of Applied Mathematics, University of Waterloo, 200 University Ave W, Waterloo, Ontario N2L 3G1, Canada}
\affiliation{Institute for Quantum Computing, University of Waterloo, 200 University Ave W, Waterloo, Ontario N2L 3G1, Canada}
\affiliation{Perimeter Institute for Theoretical Physics, 31 Caroline St. N., Waterloo, Ontario N2L 2Y5, Canada}
\affiliation{Department of Physics and Astronomy, University of Waterloo,  200 University Ave W, Waterloo, Ontario N2L 3G1, Canada}
\author{Dyuman Bhattacharya}
\email{d7bhatta@uwaterloo.ca}
\affiliation{Department of Physics and Astronomy, University of Waterloo,  200 University Ave W, Waterloo, Ontario N2L 3G1, Canada}
\author{Robert B. Mann}
\email{rbmann@uwaterloo.ca}
\affiliation{Department of Physics and Astronomy, University of Waterloo,  200 University Ave W, Waterloo, Ontario N2L 3G1, Canada}
\affiliation{Perimeter Institute for Theoretical Physics, 31 Caroline St. N., Waterloo, Ontario N2L 2Y5, Canada}
\affiliation{Department of Applied Mathematics, University of Waterloo, 200 University Ave W, Waterloo, Ontario N2L 3G1, Canada}
\affiliation{Institute for Quantum Computing, University of Waterloo, 200 University Ave W, Waterloo, Ontario N2L 3G1, Canada}

\begin{abstract}
We embed relativistic quantum information protocols in AdS/CFT: an Unruh-DeWitt detector coupled to a local primary of a holographic CFT has a reduced state fixed by the universal boundary Wightman function. We find that the mana generated in a qutrit probe reads off the boundary condition of the dual bulk scalar, de-excitation spectroscopy tracks the double-trace flow through the lowest cylinder gap, and a local boundary detector is inequivalent to the HKLL representation of a bulk-local one.
\end{abstract}

\maketitle

\tcb{\textit{Introduction}}---What can a single  finite-dimensional quantum system reveal about a strongly coupled quantum field theory, and (via holography) about a dual gravitational spacetime? Relativistic quantum information (RQI) provides the natural language for posing such questions operationally: coupling a localized probe, an Unruh-DeWitt (UDW) detector, to a quantum field yields transition probabilities, coherences, and quantum information resources from the probe's reduced state \cite{Unruh:1976db,DeWitt1979,alsing2012observer,Ralph:2012mdp,Fuentes-Schuller:2004iaz}, with the detector dynamics fixed by field correlation functions along its worldline. RQI protocols have identified pathways toward realization in quantum-optical and analog platforms \cite{Rideout:2012jb,Pozas-Kerstjens:2016rsh,zych2012general,Hu:2012jr,kent2012quantum}, but carrying them into curved spacetimes, e.g., black hole, cosmological, and anti-de Sitter settings, is demanding: closed-form Wightman functions exist only for highly symmetric geometries, analyses involve nontrivial mode sums and frequently rely on numerics \cite{Caribe:2023fhr}, and typically access limited observables such as response rates \cite{Hodgkinson:2014iua} rather than the full reduced detector state.

Moreover, a complete theory of quantum gravity remains a central open problem. The holographic principle \cite{tHooft:1993dmi,Susskind:1994vu}, most concretely realized by the anti-de Sitter/Conformal Field Theory (AdS/CFT) correspondence \cite{Maldacena:1997re,Gubser:1998bc,Witten:1998qj}, asserts that quantum gravity in $(d{+}1)$-dimensional asymptotically AdS spacetimes is equivalent to a nongravitational CFT in $d$ dimensions, namely a holographic CFT, and has delivered profound insights, including the geometric encoding of boundary entanglement and other quantum information quantities \cite{Ryu:2006bv,Myers:2010tj,Almheiri:2014lwa,Brown:2015bva,Jafferis:2015del,Belin:2021bga}. Yet many hallmark holographic observables are intrinsically nonlocal and not operational \cite{Harlow:2014yka,Takayanagi:2025ula}, leaving a gap between holographic information geometry and concrete measurement protocols.

If holography defines quantum gravity via a CFT, detector-based protocols formulated directly in the CFT furnish operational probes of holographic dynamics. Conformal symmetry fixes the vacuum two-point function of a primary to a universal form, so the reduced state of a localized probe follows from boundary data while admitting an independent bulk interpretation through the dictionary. Earlier detector-based studies in AdS and BTZ coupled the probe to the bulk field \cite{Jennings:2010vk,Henderson:2017yuv,Henderson:2018lcy}; here the detector couples directly to the boundary operator, the bulk entering only through the dictionary. We develop this boundary-first framework and extract three results: (i) the mana generated in a qutrit probe \cite{Nystrom:2024oeq}, reported as a counterterm-independent lower bound, discriminates the boundary condition, i.e.\ the quantization, of the dual bulk scalar; (ii) the double-trace flow connecting the two quantizations becomes continuous spectroscopy: a de-excitation resonance locates the lowest cylinder gap $\Delta_{\rm gap}(f)$ and inverts into a normalization-independent estimator of the flow coupling; and (iii) a local boundary coupling is operationally inequivalent to the boundary (HKLL) representation of a bulk-local detector. Throughout, the detector's mana is an interaction-dependent probe observable encoding CFT spectral data, not a certificate of pre-existing field resources, consistent with the induced-resource-theory analysis of \cite{Nystrom:2026induced}.

\tcb{\textit{Detectors in CFT}}---
We consider a $d$-dimensional CFT living on the Lorentzian cylinder $\mathcal{B}_d \cong \mathbb{R}_\tau \times S^{d-1}_R$ with metric
\begin{equation}\label{fjerioi9i}
\mathrm{d}s_b^2=-\mathrm{d}\tau^2+R^2 \mathrm{d}\Omega_{d-1}^2,
\end{equation}
where $R$ is the boundary curvature radius. We denote the CFT vacuum by $\ket{0}$, consider a scalar primary $\hat{\mathcal{O}}(x)$ of dimension $\Delta$ with $\bra{0}\hat{\mathcal{O}}\ket{0}=0$, and write $\hat{H}_{\mathrm{CFT}}$ for the generator of $\tau$-translations on $\mathcal{B}_d$.

To probe the CFT, we introduce a localized UDW detector with free Hamiltonian $\hat{H}_D$ and energy eigenstates $\{\ket{n}_D\}$ on the static worldline $x_D(\tau)=(\tau,\Omega_D)$, with $\tau$ its proper time. The dynamics is governed by the pulled-back Wightman function $\mathcal{W}(x_D,x_D^{\prime}):=\bra{0}\hat{\mathcal{O}}(x_D)\hat{\mathcal{O}}(x_D^{\prime})\ket{0}$.

The free Hamiltonian of the combined system is
$
\hat{H}_0=\hat{H}_D+\hat{H}_{\mathrm{CFT}}.
$
In the interaction picture, we take a linear coupling between the detector and the CFT operator along the worldline,
\begin{equation}\label{eq:Hint_boundary}
\hat{H}_{\mathrm{int}}(\tau)
=\lambda\,\chi(\tau)\,\hat{\mu}(\tau)\,\hat{\mathcal{O}}\big(x_D(\tau)\big).
\end{equation}
Here $\chi(\tau)$ is a real, rapidly decaying switching function, $\hat{\mu}(\tau)=e^{i\hat{H}_D\tau}\hat{\mu}(0)e^{-i\hat{H}_D\tau}$ is the detector monopole operator, and $\hat{\mathcal{O}}(x_D(\tau))$ evolves under $\hat{H}_{\mathrm{CFT}}$. Since $[\lambda]=({\rm length})^{\Delta-1}$, comparisons across different $\Delta$ are made at fixed dimensionless coupling $\tilde\lambda\equiv\lambda R^{1-\Delta}$; $\tilde\lambda$ is treated perturbatively.

Time evolution is the Dyson-ordered unitary
$
\hat{U}
=\mathcal{T}\exp[-i\int_{-\infty}^{+\infty} \mathrm{d}\tau\ \hat{H}_{\mathrm{int}}(\tau)],
$
with $\mathcal{T}$ denoting $\tau$-ordering. For an initially uncorrelated state with the detector in its ground state and the CFT in the vacuum,
$
\hat{\rho}_0=\big(\ket{0}_D\otimes \ket{0}\big)\big({}_D\bra{0}\otimes\bra{0}\big),
$
the detector's reduced density operator after the interaction is
$\hat{\rho}_D=\mathrm{Tr}_{\mathrm{CFT}}(\hat{U}\hat{\rho}_0\hat{U}^{\dagger})$,
computed order by order in $\lambda$; it encodes the excitation probabilities and coherences used below.

\tcb{\textit{Holographic setup}}---We now specialize to the standard AdS/CFT setting where the boundary CFT on $\mathcal{B}_d $ is realized as the conformal boundary of a $(d{+}1)$-dimensional AdS spacetime \cite{Aharony:1999ti}. For concreteness we take pure AdS in global coordinates, $\mathrm{d}s^2_{\mathrm{AdS}}= -h(r)\,\mathrm{d}t^2+h(r)^{-1}\mathrm{d}r^2+r^2 \mathrm{d}\Omega_{d-1}^2$ with $h(r)=1+r^2/\ell^2$, where $\ell$ is the AdS radius. A cutoff surface $r=r_c$, Weyl-rescaled with conformal factor $\omega=R/\ell=\tau/t$, reproduces the cylinder metric \eqref{fjerioi9i} as $r_c\to\infty$, fixing the boundary conformal frame and $\hat{H}_{\mathrm{CFT}}$ as the $\tau$-translation generator. Within the AdS/CFT dictionary, a scalar primary operator $\hat{\mathcal{O}}$ of dimension $\Delta$ is dual to a bulk scalar field $\Phi$, treated here as a probe on the fixed AdS background; throughout we work to leading order in the large-$N$ (generalized-free-field) expansion of the holographic CFT. Allowing a conformal curvature coupling, $\Phi$ obeys
$
\left(\Box - m^2 - \xi\,\mathcal{R}\right)\Phi=0
$
with $\mathcal{R}=-d(d+1)/\ell^2$ and $\xi=(d-1)/(4d)$. The bulk Wightman function is $\mathcal{W}_{\mathrm{bulk}}(x,x^{\prime})=\bra{0}\Phi(x)\Phi(x^{\prime})\ket{0}$.
The scaling dimension $\Delta$ is related to the effective bulk mass $m_{\mathrm{eff}}^2\equiv m^2+\xi\,\mathcal{R}$ by the standard mass-dimension relation \cite{Gubser:1998bc,Witten:1998qj,Klebanov:1999tb}
\begin{equation}
m_{\mathrm{eff}}^2\ell^2=\Delta(\Delta-d).
\label{eq:mass_dimension}
\end{equation}
For a massless bulk scalar, $m_{\mathrm{eff}}^2\ell^2=-(d^2-1)/4$, and \eqref{eq:mass_dimension} admits two roots $\Delta_\pm=\frac{1}{2}(d\pm 1)$
with $\Delta_+$ ($\Delta_-$) the standard Dirichlet-like (alternate Neumann-like) quantization. Since $m_{\mathrm{eff}}^2$ lies in the Breitenlohner-Freedman window \cite{Breitenlohner:1982jf,Breitenlohner:1982bm}, both are admissible.

\tcb{\textit{Probing the holographic CFT}}---
We now probe the holographic CFT through an operational observable: the magic monotone mana generated in a single detector. Mana quantifies the non-stabilizer resources enabling quantum computation beyond efficient classical simulation \cite{Emerson:2013zse}; the mana generated by the switched coupling thus probes the boundary operator sector.

Conformal symmetry fixes the Lorentzian Wightman function of a scalar primary $\hat{\mathcal{O}}$ of scaling dimension $\Delta$ on the cylinder $\mathcal{B}_d$ to a universal form \cite{Simmons-Duffin:2016gjk,Bak:2017rpp,OsbornPetkou1994,Ginsparg1988,Rychkov2016}
\begin{equation}
\mathcal{W}(\tau,\tau^{\prime})
=
\frac{C}{
\Big[2R^2\big(\cos\big(\tfrac{\tau-\tau^{\prime}-i\epsilon}{R}\big)-\cos\gamma\big)\Big]^{\Delta}},
\label{dijiod}
\end{equation}
where $\epsilon\to 0^+$ and the normalization
$
C=\frac{(2\Delta-d)\Gamma(\Delta)}{\pi^{d/2}\Gamma(\Delta-d/2)}
$
is fixed by holographic renormalization \cite{Freedman:1998tz} (see \cite{Miyaji:2015woj,Bak:2017rpp} for other conventions). We focus on $d=3$ and a static detector, $\gamma=0$.

We model the probe as a qutrit with computational basis
$\{\ket{0}_D,\ket{1}_D,\ket{2}_D\}$ and $\hat{H}_D=\mathrm{diag}(0,\Omega_1,\Omega_1+\Omega_2)$, taking equal adjacent gaps $\Omega_1=\Omega_2\equiv\Omega$.
In the interaction picture, the monopole moment operator is chosen as
$
\hat{\mu}(\tau)
=
\left(
\ket{1}_D\,{}_D\bra{0}\,e^{i\Omega\tau}
+\ket{2}_D\,{}_D\bra{1}\,e^{i\Omega\tau}
\right)/\sqrt{2}
+\mathrm{H.c.},
$
which generates only adjacent transitions. We choose a Gaussian switching function
$
\chi(\tau)=\exp\left(-\tau^2/(2\sigma^2)\right),
$
with interaction duration controlled by $\sigma$. We use
\begin{align}\label{Mana}
    M&=  \ln \big[1-q+\frac{1}{3}(|q-\operatorname{Re}(\beta)-\sqrt{3} \operatorname{Im}(\beta)|\nonumber\\
    &\quad+|q+2 \operatorname{Re}(\beta)|  +|q-\operatorname{Re}(\beta)+\sqrt{3} \operatorname{Im}(\beta)|)\big]
\end{align}
 to quantify the detector's mana \cite{Emerson:2013zse}, computed in the energy eigenbasis of $\hat H_D$, where the excitation probability
$q={}_D\bra{1}\hat{\rho}_D\ket{1}_D$
and the coherence
$\beta={}_D\bra{2}\hat{\rho}_D\ket{0}_D$. The remaining entries of $\hat\rho_D$ vanish (odd correlators of $\hat{\mathcal O}$ vanish at leading order in $1/N$) or first arise at $O(\lambda^4)$, required for exact positivity but shifting $M$ only at $O(\lambda^4)$. To second order in $\lambda$,
\begin{equation}
q=\frac{\lambda^2}{2} \int_{\mathbb{R}} \mathrm{d} \tau \mathrm{d} \tau^{\prime} \,
\chi(\tau) \chi\left(\tau^{\prime}\right)
e^{-i \Omega\left(\tau-\tau^{\prime}\right)} \mathcal{W}\left(\tau, \tau^{\prime}\right),
\label{eq:q_integral}
\end{equation}
while $\beta$ is the corresponding time-ordered double integral, given by eq. \eqref{eq:beta_integral} in End Matter A.

Defining $s:=\tau-\tau^{\prime}$ we rewrite $\mathcal{W}(\tau,\tau^{\prime})$ as an absolutely convergent regulated series for $\epsilon>0$, followed by the distributional limit $\epsilon\to0^+$ obtaining
\begin{align}\label{fjeio}
\mathcal{W}(s)=\frac{C}{R^{2\Delta}}\sum_{n=0}^{\infty}B_n \exp\left(-i\omega_n s\right)\exp\left(-\omega_n\epsilon\right)
\end{align}
(details are in ref. \cite{Yang:2025zrl}),
where  $\omega_n\equiv(\Delta+n)/R$ and 
$B_n\equiv\binom{2\Delta+n-1}{n}$. 
Inserting this into \eqref{eq:q_integral} yields 
\begin{equation}
q=\frac{\lambda^{2}\pi\sigma^{2}C}{R^{2\Delta}}
\sum_{n=0}^{\infty}B_n\exp\left[-\sigma^{2}(\Omega+\omega_n)^{2}\right] 
\label{eq:q_modesum}
\end{equation}
which is finite. The corresponding mode sum for $\beta$, however, contains a divergent contribution: the antisymmetric (principal-value) part of the time-ordered kernel produces an erfi series whose terms approach $(R/\sigma)\,n^{2\Delta-2}/[\sqrt{\pi}\,\Gamma(2\Delta)]$ and therefore do not sum to a finite result (End Matter A). The divergence is purely imaginary, carries the fixed envelope $e^{-\sigma^2\Omega^2}$, and is proportional to the matrix element of the local worldline operator $\chi(\tau)^2\hat\mu(\tau)^2$; it is therefore removed by the corresponding counterterm in the detector Hamiltonian. We adopt the renormalization condition $\operatorname{Im}\beta_{\rm ren}=0$ (the minimal prescription), under which $q$ and 
\begin{align}
 \beta_{\rm MS}\equiv\operatorname{Re}\beta=-\frac{\lambda^2\pi\sigma^2 C}{2R^{2\Delta}}\sum_{n=0}^{\infty}B_ne^{- \sigma^2\left(\Omega^2+\omega_n^2\right)}
 \label{eq:betaMS}
\end{align}
are counterterm independent. Every required counterterm shifts only $\operatorname{Im}\beta$; we therefore obtain the phase-minimized mana $\bar{M}\equiv\min_{\operatorname{Im}\beta}M(q,\beta)$ where
\begin{align}\label{Manamin}
\bar{M} =\ln\!\big[1-q+\tfrac13(2|q-\beta_{\rm MS}|+|q+2\beta_{\rm MS}|)\big]
\end{align}
is a counterterm-independent lower bound saturated at $\operatorname{Im}\beta_{\rm ren}=0$. Since the tower $\omega_n=(\Delta+n)/R$ shifts trivially with $\Delta$, any spectral observable separates the endpoints $\Delta_\pm$. The nontrivial content lies in the renormalized state $(q,\beta_{\rm MS})$ and the calibrated comparisons below. The two fixed points are connected by a double-trace deformation $\delta S_f=(f/2)\int \mathrm{d}^3 x\, \hat{\mathcal O}_{\Delta_-}^2$ \cite{Klebanov:1999tb,Witten:2001ua,Berkooz:2002ug,Hartman:2006dy}, which the detector resolves continuously.

\tcb{\textit{Double-trace flow}}---On the Euclidean cylinder ($\tau_E=i\tau$), the deformation drives the flow from the $\Delta_-$ theory in the ultraviolet (UV) to the $\Delta_+$ theory in the infrared (IR). At leading order in the large-$N$ expansion, the undeformed correlator $G_-$ is diagonal in Euclidean frequency $\nu$ and angular momentum $(\ell,m)$, with eigenvalue $\mathfrak g^-_\ell(\nu)$ (End Matter B). Resumming the double-trace bubbles gives \cite{Witten:2001ua,Hartman:2006dy,Berenstein:2014cia}
\begin{equation}
\mathfrak g^{(f)}_\ell(\nu)
=\frac{\mathfrak g^-_\ell(\nu)}{1+\kappa\mathfrak g^-_\ell(\nu)}
=\frac{1}{\kappa-\mathcal F_\ell(i\nu)},
\quad \kappa\equiv fR\geq0,
\label{eq:double_trace_cylinder_resummation}
\end{equation}
with
\begin{equation}
\mathcal F_\ell(x)=-2
\frac{\Gamma\!\left(\frac{\ell+2-x}{2}\right)
      \Gamma\!\left(\frac{\ell+2+x}{2}\right)}
     {\Gamma\!\left(\frac{\ell+1-x}{2}\right)
      \Gamma\!\left(\frac{\ell+1+x}{2}\right)}.
\label{eq:double_trace_robin_function}
\end{equation}
Analytic continuation yields a discrete spectrum of positive frequencies $\omega_{n\ell}=x_{n\ell}/R$ and residues $Z_{n\ell}$ determined by
\begin{equation}
\begin{aligned}
\mathcal F_\ell(x_{n\ell})&=\kappa,\quad\ell+1+2n\leq x_{n\ell}<\ell+2+2n,\\
Z_{n\ell}&=\left[\partial_x\mathcal F_\ell(x_{n\ell})\right]^{-1}.
\end{aligned}
\label{eq:double_trace_global_roots}
\end{equation}
At $\kappa=0$ this reproduces the $\Delta_-=1$ correlator \eqref{fjeio}; as $\kappa\to\infty$ the poles approach $x_{n\ell}=\ell+2+2n$ and $\kappa^2Z_{n\ell}$ the $\Delta_+=2$ weight. After removal of a contact term, $f^2W_f\to W_+$ at separated points, this rescaling reflecting the fixed UV normalization of $\hat{\mathcal O}_-$ (End Matter B).

\begin{figure}[t!]
    \centering
    \includegraphics[width=\linewidth]{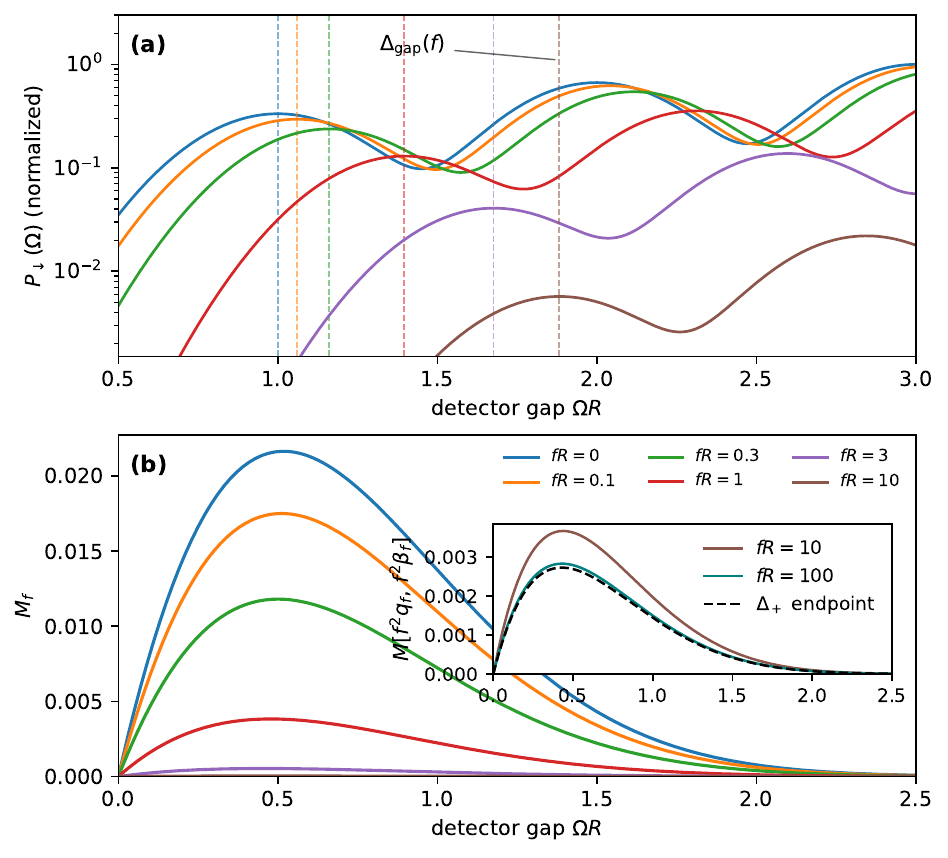}
    \caption{(a) De-excitation spectroscopy \eqref{eq:Pdown} for a detector prepared in $\ket{1}_D$, $\sigma_{\rm sp}=3R$, normalized to the $fR=0$ peak. The lowest resonance tracks $\Delta_{\rm gap}(f)$ (dashed verticals) and its height tracks $Z_{00}$. (b) Cylinder double-trace flow of the detector mana for $d=3$, $\Delta_-=1$, at fixed UV normalization ($R=\sigma=\lambda=1$, modes $x_{n\ell}\leq12$). $fR=0$ corresponds to the $\Delta_-$ endpoint. Inset: the mana of the rescaled IR operator $f\hat{\mathcal O}_-$, from $(f^2q_f,f^2\beta_f)$ at $fR=10,\,100$, collapses onto the exact $\Delta_+=2$ endpoint (dashed; deviations $34\%$ and $3.7\%$: the $O(1/fR)$ approach).}
    \label{fig:double_trace_flow}
\end{figure}

The lowest global frequency defines the dimensionless cylinder gap
    \begin{align}
\Delta_{\rm gap}(f)&\equiv x_{00}(fR)=R\,\omega_{00}(f)\nonumber,\\\mathcal F_0(\Delta_{\rm gap})&=-x_{00}\cot\left(\frac{\pi x_{00}}{2}\right)=fR,
\label{eq:Delta_eff}
    \end{align}
a finite-size gap proxy ($\Delta_{\rm gap}$ is a scaling dimension only at the fixed points) that increases monotonically from $\Delta_-=1$ to $\Delta_+=2$ along the flow. Thus, the gap is directly measurable, as a  detector prepared in $\ket{1}_D$ de-excites with probability
\begin{equation}
P_\downarrow(\Omega)=\frac{\lambda^2\pi\sigma_{\rm sp}^2}{R^2}\sum_{\ell,n}\frac{2\ell+1}{4\pi}Z_{n\ell}e^{-\sigma_{\rm sp}^2\!\left(\Omega-\frac{x_{n\ell}}{R}\right)^{\!2}},
\label{eq:Pdown}
\end{equation}
whose Gaussian-broadened resonances sit at the flow frequencies [Fig.~\ref{fig:double_trace_flow}(a)]. A switching width $\sigma_{\rm sp}\gtrsim3R$ resolves the lowest resonance, which recovers $\Delta_{\rm gap}$ to better than $2\times10^{-3}$ and inverts into the estimator $fR_{\rm est}=\mathcal F_0(R\,\Omega_{\rm peak})$, accurate to $1\%$ over $fR\in[0.1,10]$ (End Matter B); peak positions, like $\beta_{\rm MS}/q$, are independent of detector coupling and response normalization ($\kappa=fR$ presumes a fixed holographic normalization of $\hat{\mathcal O}_{\Delta_-}$). Note that we can also consider higher excitation frequencies, such as $x_{01}$ (the $\ell=1$ dipole mode). Moreover, a three-level probe allows simultaneous locking onto multiple flow resonances (e.g., $\Omega_{10}=x_{00} / R$ and $\Omega_{21}=x_{01} / R$ ). It facilitates cross-verification of the coupling $fR$ in a single operational setup, enhancing measurement precision and resilience against environmental noise.

For the static detector, $\gamma=0$, the same Gaussian kernels and the minimal condition $\operatorname{Im}\beta_{\rm ren}=0$ imposed at each $f$ give
\begin{align}
q_f&=\frac{\lambda^2\pi\sigma^2}{R^2}
\sum_{\ell,n}\frac{2\ell+1}{4\pi}Z_{n\ell}
\exp\!\left[-\sigma^2\!\left(\Omega+\frac{x_{n\ell}}{R}\!\right)^{\! 2}\!\right],
\label{eq:qf_double_trace_cylinder}\\
\beta_{f}&=-\frac{\lambda^2\pi\sigma^2}{2R^2}
e^{-\sigma^2\Omega^2}
\!\sum_{\ell,n}\!\frac{2\ell+1}{4\pi}Z_{n\ell}
\exp\!\left[-\sigma^2\frac{x_{n\ell}^2}{R^2}\right].
\label{eq:betaf_double_trace_cylinder}
\end{align}
Figure~\ref{fig:double_trace_flow}(b) evaluates the mana from \eqref{eq:qf_double_trace_cylinder}--\eqref{eq:betaf_double_trace_cylinder}: increasing $fR$ raises every global frequency from its alternate- toward its standard-quantization value while reducing its residue, and the mana decreases pointwise across the parameter ranges studied. The mana of the rescaled IR operator, from $(f^2q_f,f^2\beta_f)$, collapses onto the $\Delta_+$ endpoint (inset), verifying $f^2W_f\to W_+$ at the level of the detector observable.

\tcb{\textit{Local versus HKLL-smeared detector protocols}}---
We now make explicit the difference between a detector coupled locally to the boundary primary and a boundary protocol that represents a bulk-local detector. In global AdS, a bulk-local scalar operator is reconstructed by an HKLL smearing of boundary data \cite{Hamilton:2006az}; hence a local bulk UDW interaction is not dual to the local boundary interaction in \eqref{eq:Hint_boundary}. The distinction is already visible at the level of the two-point mode weights. Both protocols probe the same global frequency tower $\omega_{N}= (\Delta+N)/R$, but assign different angular and radial weights to its modes. Their operational distinction lies in these spectral weights, not in the locations of the frequencies themselves.

For $d=3$, the local boundary correlator decomposes into global modes,
\begin{equation}
W_{\rm loc}(s,\gamma)
=C R^{-2\Delta}\sum_{N=0}^{\infty}
 e^{-i(\Delta+N)(s-i\epsilon)/R}\, C_N^{\Delta}(\cos\gamma),
\label{eq:local_boundary_mode_decomposition}
\end{equation}
whose coincident-angle weights $A_N^{\rm loc}=\sum_{\ell}a_{N\ell}=\binom{2\Delta+N-1}{N}$ reproduce the point-detector expansion \eqref{fjeio}; here $a_{N\ell}\geq0$ are the Legendre components of the Gegenbauer polynomial $C_N^{\Delta}$, nonvanishing for $\ell=N,N-2,\ldots$ (End Matter C).

\begin{figure}[t!]
    \centering
    \includegraphics[width=\linewidth]{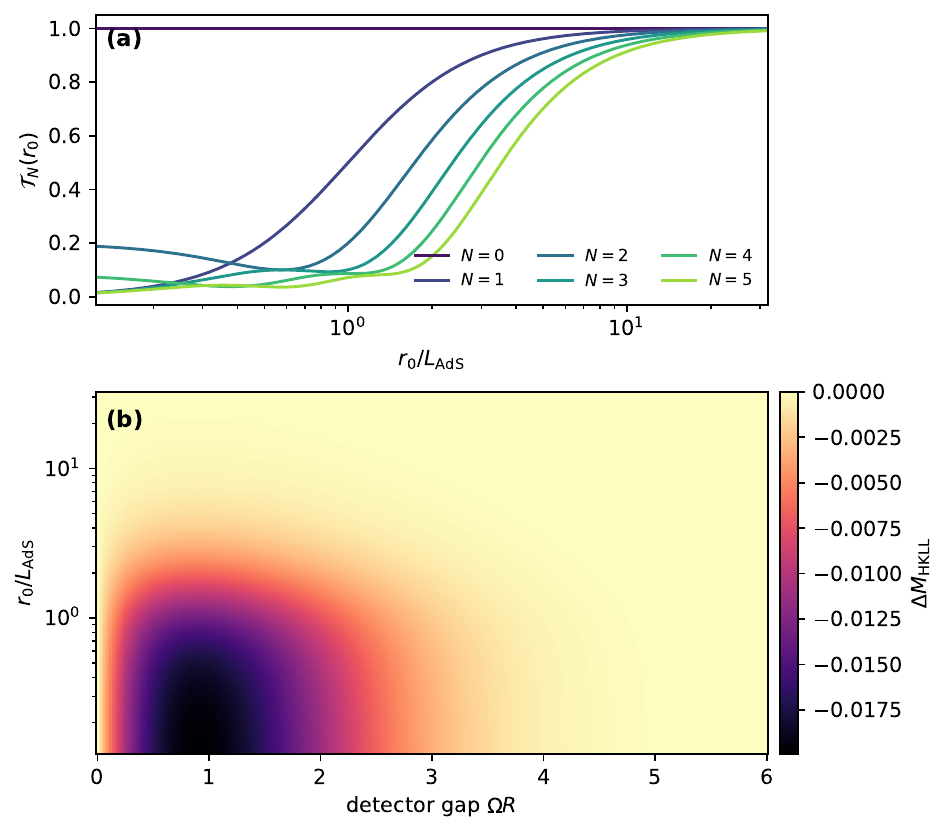}
    \caption{(a) Radial transfer function $\mathcal T_N(r_0)$ for $N\leq5$: $\mathcal T_0\equiv1$, odd-$N$ filters vanish at the center, and peak-height ratios of $P_\downarrow$ measure $\mathcal T_N$ directly. (b) Mode-sum comparison of a local boundary detector with the dimension-matched HKLL representation of a bulk-local one: mana mismatch $\Delta M_{\rm HKLL}=M_{\rm HKLL}-M_{\rm loc}$ vs detector gap and radial position $r_0/L_{\rm AdS}$ (log axis); $d=3$, $\Delta=2$, $L_{\rm AdS}=R=1$, $\lambda=1$, $\sigma/R=0.5$, $N_{\max}=80$, all quantities fixed in boundary time (End Matter C).}
    \label{fig:hkll_mode_sum_comparison}
\end{figure}

With $L_{\rm AdS}$ the AdS radius (distinguished here from the angular-momentum label $\ell$) and bulk static time obeying $\tau=(R/L_{\rm AdS})t$, the bulk-local protocol is defined by the dimension-matched HKLL operator
\begin{equation}
\begin{aligned}
\widehat{\mathcal O}_{r_0}^{\rm HKLL}(\tau,\Omega_D)
\equiv&\left(1+\frac{r_0^2}{L_{\rm AdS}^2}\right)^{\Delta/2}\\
&\times\Phi\!\left(t=\frac{L_{\rm AdS}}{R}\tau,r_0,\Omega_D\right).
\end{aligned}
\label{eq:hkll_operator_tr}
\end{equation}
Expanded in the global AdS$_4$ modes---whose frequencies $\omega_{n\ell}L_{\rm AdS}=\Delta+2n+\ell$ match the boundary tower with $N=2n+\ell$ (End Matter C)---this operator acts in mode space as a radial filter $\mathcal K_{N\ell}(r_0)$ on each Legendre component, so the correlator sampled by the detector is
\begin{equation}
\begin{aligned}
W_{\rm HKLL}(s;r_0)
&=C R^{-2\Delta}\sum_{N=0}^{\infty}A_N^{\rm HKLL}(r_0)
 e^{-i(\Delta+N)(s-i\epsilon)/R},\\
A_N^{\rm HKLL}(r_0)
&=\sum_{\ell}a_{N\ell}\,|\mathcal{K}_{N\ell}(r_0)|^2.
\end{aligned}
\label{eq:hkll_weight}
\end{equation}
End Matter C gives the Jacobi-polynomial form of $\mathcal K_{N\ell}$. As $r_0\to\infty$ at fixed $(N,\ell)$, $\mathcal K_{N\ell}\to1$ and the local protocol is recovered at the asymptotic boundary; at the AdS center only the $s$-wave, even-level tower survives, $|\mathcal K_{N\ell}(0)|=\delta_{\ell 0}$ for $\Delta=2$ (odd $N$ absent). Since $a_{N\ell}\geq0$ and, numerically, $|\mathcal K_{N\ell}|\leq1$ for all $N\leq80$ explored (End Matter C), a finite bulk radius only reduces mode weights in this range; the protocols differ through the calibration-independent radial transfer function $\mathcal T_N(r_0)\equiv A^{\rm HKLL}_N(r_0)/A^{\rm loc}_N\in[0,1]$ [Fig.~\ref{fig:hkll_mode_sum_comparison}(a)]. De-excitation peak-height ratios measure $\mathcal T_N$ directly and invert to $r_0$ (End Matter C).

Applying the Gaussian kernels and the minimal condition to each protocol yields $(q_X,\beta_{{\rm min},X})$ for $X={\rm loc},{\rm HKLL}$ (End Matter C). The two worldline correlators have different coincidence-limit structures, so each carries its own counterterm, and the comparison below is a protocol comparison at fixed detector calibration rather than a scheme-independent ordering (End Matter A). The calibration is fixed in \emph{boundary} time: varying $r_0$ at fixed $(\Omega,\sigma,\lambda)$ changes the proper gap, duration, and coupling through the redshift $\sqrt{h(r_0)}$, so Fig.~\ref{fig:hkll_mode_sum_comparison} compares boundary-time protocols, not one fixed proper-time apparatus (conversions in End Matter C). Figure~\ref{fig:hkll_mode_sum_comparison}(b) plots the mana mismatch
\begin{equation}
\Delta M_{\rm HKLL}(r_0,\Omega)=M_{\rm HKLL}(r_0,\Omega)-M_{\rm loc}(\Omega)
\label{eq:mana_mismatch}
\end{equation}
for the parameters of Fig.~\ref{fig:hkll_mode_sum_comparison}; the short switching $\sigma/R=0.5$ exposes the higher modes. The mismatch is negative: the filter suppresses descendant weights, so at fixed calibration the bulk-interior protocol generates less coherence and, in the plotted range, lower mana. A local boundary detector and the HKLL representation of a bulk-local detector are thus different operational probes of the same primary sector.

\tcb{\textit{Discussion}}---
We have developed a boundary-first formulation of RQI protocols in holography: a localized UDW detector coupled to a scalar primary, whose reduced state follows perturbatively from the universal CFT two-point function. Unlike intrinsically nonlocal holographic observables, the detector state encodes laboratory-style readouts, transition probabilities and coherences, sensitive to bulk data, the dimension $\Delta$ is fixed by the bulk quantization choice and read off by spectroscopy. Concretely, the mana generated in a qutrit probe---a counterterm-independent lower bound---distinguishes the admissible quantizations; the double-trace analysis upgrades this into continuous spectroscopy as $\Delta_{\rm gap}(f)$ is located by a de-excitation resonance and inverted into an estimator of $fR$, with the rescaled IR output collapsing onto the standard-quantization endpoint; and the HKLL comparison shows a local boundary coupling and the boundary representation of a bulk-local coupling are not interchangeable, the difference is the transfer function $\mathcal T_N(r_0)$, which the same de-excitation protocol measures through peak-height ratios, reconstructing the bulk radius $r_0$ operationally. The endpoint quantizations also differ by the classic double-trace change in the sphere free energy \cite{Gubser:2002vv,Klebanov:2011gs}, consistent with the $F$-theorem; accessing $\delta F$ operationally would require stress-tensor couplings. Since UDW detectors admit quantum-optical and analog realizations \cite{Rideout:2012jb,Pozas-Kerstjens:2016rsh} and mana is the magic-state-distillation resource \cite{Emerson:2013zse}, these observables tie holographic data to operationally meaningful quantities.

The same strategy applies whenever a holographic dictionary supplies controlled boundary correlators: dS/CFT \cite{Strominger:2001pn,Maldacena:2002vr,Spradlin:2001pw,Anninos:2011ui}, with the Lorentzian input fixed by the analytic continuation and $i\epsilon$ prescription, including complex $\Delta\simeq d/2\pm im\ell$; BTZ/CFT \cite{Keski-Vakkuri:1998gmz}; defect and boundary CFT \cite{Takayanagi:2011zk,Bissi:2018mcq,Mazac:2018biw,Hao:2025ocu}; nonunitary \cite{Brody:2013axr,Chang:2019jcj} and complex \cite{Doi:2022iyj,Jacobsen:2024jel,Tang:2024blm} CFTs; Carrollian \cite{Barnich:2010eb} and celestial \cite{Pasterski:2016qvg} holography. Besides, integrating our boundary detector framework with measurement-induced state update protocols could offer deeper insights into how information gain and measurement disturbance map onto bulk quantum geometries \cite{Balasubramanian:2025qoz}. On the RQI side, entanglement harvesting, communication capacities, metrology, and channel discrimination import directly with the same machinery \cite{Avalos:2022oxr}; on the holography side, detector-based resources probe operator sectors, quantization choices, deformations, and bulk reconstruction. Detector-based probes thus provide a calculable bridge between holographic CFT data and experimentally motivated quantum-information protocols.

\textit{Acknowledgments}\kern0pt---We are grateful to Rob Myers for presenting the work of Ref.~\cite{Wurtz:2026ehc} at the conference ``Gravity, Particles, and Information: Celebrating Robert B. Mann's Contributions to Physics'' and for helpful discussions thereafter. We extend our thanks to William Unruh for his illuminating talk and discussions. We are also grateful to Vijay Balasubramanian, Gong Cheng, Thomas Hartman, Hong Lu, Nicola Pranzini, Edward Witten, Xi Yin, and Jiaju Zhang for valuable discussions and insightful comments. This work was supported by the Natural Sciences and Engineering Research Council of Canada, the National Natural Science Foundation of China (Grant No. 12365010), and the China Scholarship Council Scholarship. Research at Perimeter Institute is supported in part by the Government of Canada through the Department of Innovation, Science and Economic Development and by the Province of Ontario through the Ministry of Colleges and Universities.

\textit{Note added}---This paper (v1) appeared on arXiv simultaneously with the independent parallel study of Wurtz, Lima, Myers, and Mart\'in-Mart\'inez \cite{Wurtz:2026ehc}, which analyzes two-detector entanglement harvesting in general $d$-dimensional CFTs. The two works are complementary: Ref.~\cite{Wurtz:2026ehc} focuses on bipartite entanglement and its separation into field-harvested and communication-mediated contributions, whereas we study single-detector mana generation, the double-trace flow, and the boundary-local vs bulk-local (HKLL) comparison.

\newpage
\mbox{}
\newpage

\section*{End Matter}

\setcounter{equation}{0}
\renewcommand{\theequation}{A\arabic{equation}}
\renewcommand{\theHequation}{A.\arabic{equation}}
\section*{Appendix A: Renormalization of the detector coherence}
To second order in $\lambda$, the $\ket{0}_D\!\to\!\ket{2}_D$ coherence is
\begin{equation}
\begin{aligned}
\beta= & -\frac{\lambda^2}{4} \int_\mathbb{R} \mathrm{d} \tau \mathrm{d} \tau^{\prime} \chi(\tau) \chi\left(\tau^{\prime}\right) \\
& \times\left[\Theta\left(\tau-\tau^{\prime}\right) e^{i\left(\Omega_1 \tau^{\prime}+\Omega_2 \tau\right)} \mathcal{W}\left(\tau, \tau^{\prime}\right)\right. \\
& \left.+\Theta\left(\tau^{\prime}-\tau\right) e^{i\left(\Omega_1 \tau+\Omega_2 \tau^{\prime}\right)} \mathcal{W}\left(\tau^{\prime}, \tau\right)\right],
\end{aligned}
\label{eq:beta_integral}
\end{equation}
where $\Theta$ is the Heaviside step function. Writing $\Theta(\tau-\tau')=\tfrac12[1+\operatorname{sgn}(\tau-\tau')]$ splits the time-ordered kernel into a symmetric part and an antisymmetric (principal-value) part. The symmetric part yields the real, absolutely convergent series $\beta_{\rm MS}$ of \eqref{eq:betaMS}. The principal-value part gives
\begin{equation}
\beta=\beta_{\rm MS}+i\,\frac{\lambda^2\pi\sigma^2C}{2R^{2\Delta}}\,
e^{-\sigma^2\Omega^2}\sum_{n=0}^{\infty}B_n\,
e^{-\sigma^2\omega_n^2}\operatorname{erfi}(\sigma\omega_n).
\label{eq:beta_split}
\end{equation}
Using $\operatorname{erfi}(z)\sim e^{z^2}/(\sqrt{\pi}z)$ and $B_n\sim n^{2\Delta-1}/\Gamma(2\Delta)$, the terms of the series in \eqref{eq:beta_split} approach $(R/\sigma)\,n^{2\Delta-2}/[\sqrt{\pi}\,\Gamma(2\Delta)]$. The divergence is thus purely imaginary: it is the dispersive part of the second-order detector evolution, sourced by the coincidence limit of the worldline correlator. Throughout, every Wightman mode sum carries the prescription $s\to s-i\epsilon$ of Eq.~\eqref{fjeio}. The Gaussian switching integrals converge absolutely mode by mode, so $\epsilon\to0^+$ commutes with all convergent sums, the principal-value series above being the sole exception.

This structure identifies the required counterterm. For the qutrit monopole operator of the main text,
\begin{equation}
\hat\mu(\tau)^2=\tfrac12\big(\ket{0}\!\bra{0}+2\ket{1}\!\bra{1}+\ket{2}\!\bra{2}\big)
+\tfrac12\big(e^{2i\Omega\tau}\ket{2}\!\bra{0}+{\rm H.c.}\big),
\label{eq:musq}
\end{equation}
so the local worldline counterterm
\begin{equation}
\delta\hat H_{\rm ct}(\tau)=\lambda^2\,c\,\chi(\tau)^2\,\hat\mu(\tau)^2,
\qquad c\in\mathbb{R}
\label{eq:counterterm}
\end{equation}
contributes at first order, $\delta\hat\rho_D=-i\int\mathrm{d}\tau\,[\delta\hat H_{\rm ct},\hat\rho_0]$, giving
\begin{equation}
\delta\beta=-\tfrac{i}{2}\lambda^2 c\sqrt{\pi}\,\sigma\, e^{-\sigma^2\Omega^2},
\qquad \delta q=\delta\rho_{00}=\delta\rho_{22}=0.
\label{eq:ct_shift}
\end{equation}
The shift is purely imaginary, with exactly the envelope of the divergence in \eqref{eq:beta_split}, and it leaves $q$ and $\operatorname{Re}\beta$ untouched. Choosing the coefficient $c$ to cancel the principal-value series renormalizes the coherence. Writing $y=\operatorname{Im}\beta$, the mana obeys $|q-\operatorname{Re}\beta-\sqrt3\,y|+|q-\operatorname{Re}\beta+\sqrt3\,y|\geq2|q-\operatorname{Re}\beta|$, so $M(q,\operatorname{Re}\beta,y)\geq M(q,\operatorname{Re}\beta,0)$ with equality iff $y=0$. The reported quantity is the minimum of the mana over the scheme phase---a counterterm-independent lower bound saturated at $\operatorname{Im}\beta_{\rm ren}=0$ (the minimal prescription). Any finite counterterm induces precisely phase freedom removed by the minimization.

Along the flow the divergences have a two-scale structure. The leading coincidence singularity of $W_f$ coincides with that of $W_-$ (the deformation is relevant), so the leading divergence of the principal-value sum is $f$-independent. However, the subleading $O(f/s)$ coincidence singularity of $W_f$ produces an $f$-\emph{dependent} logarithmic divergence. No single $f$-independent subtraction therefore renormalizes the entire flow. The robustness of the reported mana is unaffected, following from the elimination of the coherence phase rather than from any universal subtraction. By contrast, the boundary-local and HKLL worldline correlators have different coincidence-limit structures, so each protocol carries its own counterterm; the comparison in Fig.~\ref{fig:hkll_mode_sum_comparison} is made with both protocols renormalized by the same condition $\operatorname{Im}\beta_{\rm ren}=0$, i.e.\ at fixed detector calibration, and is not asserted as a scheme-independent ordering.

\setcounter{equation}{0}
\renewcommand{\theequation}{B\arabic{equation}}
\renewcommand{\theHequation}{B.\arabic{equation}}
\section*{Appendix B: Double-trace spectral data and numerical checks}
With $u=(\tau_E-\tau_E')/R$ and $\mathbf n,\mathbf n'\in S^2$, the undeformed cylinder correlator and its angular-frequency decomposition are
\begin{equation}
\begin{aligned}
G_-(u,\mathbf n,\mathbf n')
&=\frac{C_-}{2R^2(\cosh u-\mathbf n\!\cdot\!\mathbf n')} \\
&=\frac{1}{R^2}\sum_{\ell m}\int_{-\infty}^{\infty}\frac{\mathrm{d}\nu}{2\pi}
 e^{i\nu u}Y_{\ell m}(\mathbf n)Y^*_{\ell m}(\mathbf n')
 \mathfrak g^-_\ell(\nu),
\end{aligned}
\label{eq:double_trace_cylinder_decomposition}
\end{equation}
where $C_-=1/(2\pi^2)$ and
\begin{equation}
\mathfrak g^-_\ell(\nu)=\frac{1}{2}
\frac{\Gamma\!\left(\frac{\ell+1+i\nu}{2}\right)
      \Gamma\!\left(\frac{\ell+1-i\nu}{2}\right)}
     {\Gamma\!\left(\frac{\ell+2+i\nu}{2}\right)
      \Gamma\!\left(\frac{\ell+2-i\nu}{2}\right)}.
\label{eq:double_trace_cylinder_eigenvalue}
\end{equation}
Analytic continuation of \eqref{eq:double_trace_cylinder_resummation} yields the deformed Wightman function
\begin{equation}
W_f(s,\gamma)=\frac{1}{R^2}\sum_{\ell=0}^{\infty}
\frac{2\ell+1}{4\pi}P_\ell(\cos\gamma)\sum_{n=0}^{\infty}Z_{n\ell}(\kappa)
 e^{-\frac{ix_{n\ell}(s-i\epsilon)}{R}},
\label{eq:double_trace_cylinder_wightman}
\end{equation}
with $P_\ell$ the Legendre polynomials. The endpoint residues are
\begin{align}
Z^{(-)}_{n\ell}&=\frac{(2n)!\,\Gamma(\ell+n+1)}{4^n(n!)^2\sqrt{\pi}\,\Gamma(\ell+n+\tfrac32)},
\label{eq:Zminus}\\
Z^{(+)}_{n\ell}&=\lim_{\kappa\to\infty}\kappa^2 Z_{n\ell}(\kappa)
=\frac{(2n+2)!\,\Gamma(\ell+n+2)}{4^n\,n!\,(n+1)!\,\sqrt{\pi}\,\Gamma(\ell+n+\tfrac32)}.
\label{eq:Zplus}
\end{align}
Summing the $(\ell,n)$ sectors at fixed descendant level reproduces the $\Delta_\mp$ cylinder coefficients $C_\mp B_n$, and at $\kappa=0$ the two-index detector sums \eqref{eq:qf_double_trace_cylinder}--\eqref{eq:betaf_double_trace_cylinder} agree with the one-index expansions \eqref{eq:q_modesum} and \eqref{eq:betaMS}.

\emph{De-excitation spectroscopy}---At second order, the probability for a detector prepared in $\ket{1}_D$ to be found in $\ket{0}_D$ is the de-excitation counterpart of Eq.~\eqref{eq:q_integral}: with $\bra{0}\hat\mu\ket{1}=1$ and switching width $\sigma_{\rm sp}$, $P_\downarrow=\lambda^2\!\int\!\mathrm{d}\tau\,\mathrm{d}\tau'\chi(\tau)\chi(\tau')\,e^{i\Omega(\tau-\tau')}W_f(\tau-\tau',0)$. The detector releases rather than absorbs the gap energy, so the anti-resonant kernels $e^{-\sigma^2(\Omega+x_{n\ell}/R)^2}$ of $q_f$ become the resonant kernels of Eq.~\eqref{eq:Pdown}; equivalently, $P_\downarrow(\Omega)=q_f(-\Omega)$ at $\sigma=\sigma_{\rm sp}$. The $1\to0$ channel is exact at this order, while the $1\to2$ channel is the nonresonant analogue of \eqref{eq:qf_double_trace_cylinder} and is negligible near the resonances. The gap above the lowest frequency obeys $x_{10}-x_{00}\geq0.91$ for all $\kappa\in[10^{-2},10^{2}]$, so the lowest resonance is resolved for $\sigma_{\rm sp}\gtrsim3R$; because the level weights grow with frequency, the operational marker is the \emph{first local maximum} of $P_\downarrow(\Omega)$, not its global maximum. Along the flow, $\Delta_{\rm gap}=1,\,1.06,\,1.16,\,1.40,\,1.68,\,1.88$ at $fR=0,\,0.1,\,0.3,\,1,\,3,\,10$. At $\sigma_{\rm sp}=3R$ the first peak deviates from $\Delta_{\rm gap}$ by at most $2\times10^{-3}$ and the inverted estimator $\mathcal F_0(R\,\Omega_{\rm peak})$ reproduces $fR$ to better than $1\%$ for $fR\in[0.1,10]$; at $\sigma_{\rm sp}=4R$ these improve to $10^{-4}$ and $0.2\%$.

\setcounter{equation}{0}
\renewcommand{\theequation}{C\arabic{equation}}
\renewcommand{\theHequation}{C.\arabic{equation}}
\section*{Appendix C: HKLL mode data}
The Legendre components of the Gegenbauer weight in \eqref{eq:local_boundary_mode_decomposition}, $C_N^{\Delta}(\cos\gamma)=\sum_{\ell=0}^N a_{N\ell}P_\ell(\cos\gamma)$, are
\begin{equation}
a_{N\ell}=\frac{2\ell+1}{2}\int_{-1}^{1}dx\,
C_N^{\Delta}(x)P_{\ell}(x)\geq0.
\label{eq:aNell_projection}
\end{equation}
The global AdS$_4$ scalar modes are $u_{n\ell m}\propto e^{-i\omega_{n\ell}t}e_{n\ell}(r)Y_{\ell m}(\Omega)$ with $\omega_{n\ell}L_{\rm AdS}=\Delta+2n+\ell$ and radial profiles
\begin{equation}
\begin{aligned}
e_{n\ell}(r)\propto&
\left(\frac{r}{L_{\rm AdS}}\right)^{\ell}
\left(1+\frac{r^2}{L_{\rm AdS}^2}\right)^{-(\Delta+\ell)/2}\\
&\times P_n^{(\ell+1/2,\Delta-3/2)}
\left(\frac{L_{\rm AdS}^2-r^2}{L_{\rm AdS}^2+r^2}\right),
\end{aligned}
\label{eq:global_ads_radial_modes}
\end{equation}
where $N=2n+\ell$, so that for $s=\tau-\tau'$ the bulk phase satisfies $e^{-i\omega_{n\ell}(t-t')}=e^{-i(\Delta+N)(s-i\epsilon)/R}$, matching the boundary spectral decomposition. The radial filter of \eqref{eq:hkll_weight} is
\begin{equation}
\begin{aligned}
\mathcal{K}_{N\ell}(r_0)=&
\left(\frac{r_0}{\sqrt{L_{\rm AdS}^2+r_0^2}}\right)^{\ell}\\
&\times\frac{P_n^{(\ell+1/2,\Delta-3/2)}
\!\left(\frac{L_{\rm AdS}^2-r_0^2}
{L_{\rm AdS}^2+r_0^2}\right)}
{P_n^{(\ell+1/2,\Delta-3/2)}(-1)},\qquad
n=\frac{N-\ell}{2}.
\end{aligned}
\label{eq:hkll_radial_filter}
\end{equation}
As $r_0\to\infty$ at fixed $(N,\ell)$, $\mathcal K_{N\ell}\to1$. At $r_0=0$ the prefactor removes all $\ell\geq1$, and for $\Delta=2$ the surviving $\ell=0$ ratio has unit magnitude, $|P_n^{(1/2,1/2)}(1)/P_n^{(1/2,1/2)}(-1)|=1$: only the $s$-wave, even-$N$ tower contributes, with unit weight. Numerically we find $|\mathcal K_{N\ell}(r_0)|\leq1$ throughout the range explored ($N\leq80$, all plotted radii), with the bound saturated only by $\mathcal K_{00}\equiv1$ (largest nontrivial value $0.9995$); we do not have an analytic proof for all $(N,\ell,r_0)$, and the main-text statement is restricted accordingly. Within this range $A^{\rm HKLL}_N(r_0)\leq A^{\rm loc}_N$ for all $N$: the filter only suppresses.

\emph{Operational reconstruction}---The ratio of de-excitation peak heights of the HKLL and local protocols at $\Omega=(\Delta+N)/R$ measures $\mathcal T_N$ directly: at $\sigma_{\rm sp}=3R$ the ratio agrees with $\mathcal T_N$ to $\leq10^{-4}$ for $N\leq4$ (inter-mode leakage $\sim e^{-\sigma_{\rm sp}^2}$). Since $\mathcal T_1(r_0)$ is monotone, the measured ratio inverts to the bulk radius: in tests at $r_0/L_{\rm AdS}=0.5,\,1,\,2$ the reconstruction error is $\leq0.02\%$. With the flow estimator $fR_{\rm est}$ (End Matter B), de-excitation spectroscopy thus reconstructs both the deformation parameter and the bulk position from detector data.

The Gaussian detector kernels then give, for $X={\rm loc}$ or ${\rm HKLL}$,
\begin{equation}
q_X=\lambda^2\pi\sigma^2 C R^{-2\Delta}
\sum_N A_N^X
\exp\left[-\sigma^2\left(\Omega+\frac{\Delta+N}{R}\right)^2\right],
\label{eq:q_hkll_local_modesum}
\end{equation}
\begin{equation}
\begin{aligned}
\beta_{{\rm min},X}=&-\frac{\lambda^2\pi\sigma^2}{2}
 e^{-\sigma^2\Omega^2} C R^{-2\Delta}\\
&\times\sum_N A_N^X
\exp\left[-\sigma^2\left(\frac{\Delta+N}{R}\right)^2\right],
\end{aligned}
\label{eq:beta_hkll_local_modesum}
\end{equation}
with each protocol renormalized by its own minimal condition (End Matter A). At $\sigma/R=1/2$, the modes neglected beyond $N_{\max}=80$ enter with weight below $e^{-\sigma^2(\Delta+N)^2/R^2}\big|_{N=81}\approx e^{-1722}$, so Fig.~\ref{fig:hkll_mode_sum_comparison} is fully convergent.

\emph{Proper-time calibration}---Along the static worldline, $\mathrm{d}\tau_p=\sqrt{h(r_0)}\,\mathrm{d}t$ with $h=1+r_0^2/L_{\rm AdS}^2$, while the boundary conformal frame uses $\mathrm{d}\tau=(R/L_{\rm AdS})\,\mathrm{d}t$, so $\mathrm{d}\tau_p=(L_{\rm AdS}/R)\sqrt{h}\,\mathrm{d}\tau$. A fixed proper apparatus $(\Omega_p,\sigma_p,\lambda_p)$ therefore corresponds to the boundary-frame values $\Omega=(L_{\rm AdS}/R)\sqrt{h}\,\Omega_p$, $\sigma=(R/L_{\rm AdS})\,\sigma_p/\sqrt{h}$, $\lambda=(L_{\rm AdS}/R)\sqrt{h}\,\lambda_p$, and the global tower appears at proper frequencies $\omega^p_N=(\Delta+N)/(L_{\rm AdS}\sqrt{h})$. Fig.~\ref{fig:hkll_mode_sum_comparison} holds $(\Omega,\sigma,\lambda)$ fixed in boundary time and thus compares boundary-calibrated protocols. A fixed proper apparatus coupled to $\Phi$ itself carries in addition the per-mode envelope $h^{-\Delta}(r_0)$, yet its response does \emph{not} decouple at large $r_0$: the number of tower modes inside the Gaussian window grows as $h^{\Delta}$ (level weights $\sim N^{2\Delta-1}$ against $\omega^p_N\propto1/\sqrt h$), compensating the envelope. At $(\Omega_p,\sigma_p)=(1,0.5)$ and $(2.5,0.5)$ in AdS units the fixed-proper mana varies by $\lesssim2\%$ over $r_0\in[0.05,5]\,L_{\rm AdS}$ (converged at $N_{\max}=80$): the strong radial dependence of Fig.~\ref{fig:hkll_mode_sum_comparison}(b) is a property of the boundary-calibrated comparison, while the calibration-invariant content is the transfer function $\mathcal T_N(r_0)=\sum_\ell a_{N\ell}|\mathcal K_{N\ell}|^2/\sum_\ell a_{N\ell}\in[0,1]$, common to all worldline parametrizations [Fig.~\ref{fig:hkll_mode_sum_comparison}(a)].

\end{document}